\documentclass[prl,amssymb,twocolumn,showpacs,superscriptaddress]{revtex4}

\usepackage{graphicx}
\usepackage{bm}

\begin{document}

\title{Critical currents at the Bragg glass to vortex glass
transition}

\author{Alexander D. Hern\'{a}ndez}
\affiliation{Laboratorio de
Superconductividad, Facultad de F\'{\i}sica-IMRE, Universidad de la
Habana, 10400, Ciudad Habana, Cuba.}
\affiliation{Centro At{\'{o}}mico Bariloche and Instituto Balseiro, 
8400 San Carlos de Bariloche,
R\'{\i}o Negro, Argentina.}
\author{Daniel Dom\'{\i}nguez}
\affiliation{Centro At{\'{o}}mico Bariloche and Instituto Balseiro, 
8400 San Carlos de Bariloche,
R\'{\i}o Negro, Argentina.}

\begin{abstract}
We present  simulations of the transport properties of superconductors
at the transition from the Bragg glass (BG) 
to  the vortex glass (VG) phase. 
We study the frustrated anisotropic 3D XY
model with point disorder, which has been shown to have a first order 
transition as a function of the intensity of disorder. 
We add an external current to the
model and  we obtain current-voltage curves as a function of disorder at
a low temperature.  We find that the in-plane critical current has
a steep increase at the BG-VG transition, while the $c$-axis critical current 
has a discontinous jump down, this later result 
in agreement with the first-order character of 
the transition.

\end{abstract}

\pacs{74.25.Qt, 74.25.Sv}

\maketitle

The study of the vortex phase diagram of anisotropic superconductors
with point disorder has been of  interest 
since the discovery of high $T_c$ superconductivity.
It is now clear that
at low temperatures and low magnetic
fields there is a ``Bragg glass'' phase (BG) with an elastically 
distorted vortex lattice \cite{giamarchi}. 
This vortex lattice undergoes a first order melting
transition  to a vortex liquid (VL) when
increasing the temperature $T$.
 At higher magnetic fields and low $T$
there is a disordered vortex state, 
the ``vortex glass'' (VG) \cite{vglass,bokil,kosterlitz,olsson}. 
A disorder driven transition from the BG to the VG has been
proposed \cite{giamarchi}
to occur when  increasing the magnetic field $H$ or the disorder.
Experimental observations  with increasing $H$ at low $T$,
that have been attributed
to the BG-VG transition, are: the destruction of the Bragg peaks 
in neutron scattering
\cite{neutron}, a dip in the differential resistance \cite{safar}, 
the onset of the ``second magnetization peak''
\cite{khay,nishizaki}, 
or a jump in the Josephson plasma resonance \cite{gaifu}. 
 Numerical simulations in particle-like models with random pinning 
have found a transition from an ordered lattice to a 
disordered lattice
\cite{ryu,zimanyi} when increasing particle density 
({\it i.e.}, the magnetic field).
Recently, Monte Carlo simulations in 
the  frustrated 3D XY model with disorder have stablished
that the BG-VG transition 
is of first-order type \cite{hu,teitel}.

Several of the experimental evidences of the BG-VG transition 
are directly or indirectly related to transport properties. 
For example, the most common
determination of the BG-VG line is through the onset of a ``second
magnetization peak'' \cite{khay,nishizaki}, which is attributed
to an steep increase of the in-plane critical currents. 
From the point of view of transport, the BG 
has zero linear resistivity ($\rho_{\rm lin}=0$)  
\cite{giamarchi,kosterlitz}.
The VG phase was originally proposed to have
zero resistivity \cite{vglass}. 
However, studies of the XY gauge glass
model with finite screening ($\kappa<\infty$) 
have found that VG is unstable  in $d=3$ \cite{bokil,kosterlitz}, 
and the disordered phase 
is a frozen vortex liquid that has a 
very small finite resistivity. 
Only for the $\kappa\rightarrow\infty$
disordered XY and gauge glass models  
the VG phase is stable
in $d=3$  at low $T$, in which
case $\rho_{\rm lin}=0$ \cite{olsson}. 
Therefore, it is important to understand how the 
the non-linear 
current-voltage (IV) curves and their critical currents
change across a disorder driven {\it first-order} transition.
We present here numerical calculations of
both the in-plane and the $c$-axis IV curves across the
{\it disorder driven} BG-VG transition 
in the $\kappa=\infty$ frustrated XY model used in \cite{hu,teitel}
(which is the only model where the first-order nature of the transition
has been clearly determined so far).
We find that the $c$-axis 
critical current has a discontinuous jump at the BG-VG transition,
 which could be  tested experimentally.

The hamiltonian of the frustrated 3D  XY model is:
$$
{H} = \sum_{{\bf r}} \sum_{{\bf \mu }=
{\bf \hat{x}},{\bf \hat{y}}} J_{{\bf r} \mu }
V(\theta _{\mu }({\bf r})) - 
\frac{J}{\Gamma}\cos({\theta _{{\bf\hat{z}}}({\bf r})})
$$
where  $\theta _{\mu }({\bf r})=
\theta({\bf r}+{\bf \mu})-\theta({\bf r})-A_{\mu}({\bf r}) $
and 
$A_{\mu}({\bf r})=\frac{2\pi}{\Phi_0} 
\int_{{\bf r}a}^{({\bf r}+{\bf\mu})a}{\bf A}\cdot d{\bf l} $
and ${\bf r}=(n_x,n_y,n_z)$ defines a lattice site in a cubic grid.
To minimize the pinning of the numerical grid, we have chosen 
$ 
V(\theta _{\mu }({\bf r})) = -R_{0} - R_{1} \cos(\theta _{\mu }({\bf r}))
-R_{2} \cos(2 \theta _{\mu }({\bf r})) $
and adjusted the coefficients $R_{0}, R_{1}$ and $R_{2}$ as in
\cite{koshelev}.
We model uncorrelated random point pinning in the xy plane
\cite{hu,teitel}
with $
J_{{\bf r} \mu }=
J(1+p \epsilon_{r, \mu})$,  ${\bf \mu }={\bf \hat{x}},{\bf
\hat{y}}$,
where $\epsilon _{r, \mu}$ are independent random variables with 
$\langle \epsilon _{r, \mu} \rangle =0$ and 
$\langle \epsilon^2 _{r, \mu} \rangle =1$.
In the presence of an external magnetic field $H$ along the $\hat z$
direction, 
we have $A_x({\bf r})-A_x({\bf r}+{\bf y})+ 
A_y({\bf r}+{\bf x})-A_y({\bf r})=2\pi f $, and we take
$f=H a^2/\Phi_0=\frac{1}{24}$.
We model the current between two grid points with the RSJ model
\cite{dd}:
$$
I_{\mu}({\bf r},t)= S_{\mu}({\bf r},t) + 
                  N_{\mu}({\bf r},t) +
		  \eta_{\mu}({\bf r},t) $$
with the superconducting current
$S_{\mu}({\bf r},t)= \frac{2e}{\hbar} 
\frac{\partial {H}}
{ \partial \theta_{\mu }({\bf r})}$, the normal current
$N_{\mu}({\bf r},t)= 
   \frac{\hbar}{2e R_{\mu}} 
\frac{\partial \theta_{\mu}({\bf r})}{\partial t}$
   with $R_{\mu}=R_{ab}$ for ${\bf \mu}={\bf\hat{x}},{\bf \hat{y}}$ 
   and $R_\mu=\Gamma R_{ab}$ for ${\bf \mu}={\bf \hat{z}}$,
and the thermal noise fluctuations have correlations 
$ \langle \eta_{\mu}({\bf r},t)\eta_{\mu '} ({\bf r'},t') \rangle = 
(2k_BT/R_\mu) \delta_{{\hat \mu},{\hat \mu '}} 
\delta_{{\bf r},{\bf r'}} \delta(t-t')$.
We take periodic boundary conditions in all directions
with a fluctuating twist $\alpha_\mu$ such that
$A_{\mu}({\bf r},t)=A_{\mu}^0({\bf r})-\alpha_{\mu}(t)$.
This allows to obtain the voltage in each direction as
$V_\mu=\frac{\hbar}{2e}\frac{ d\alpha_\mu(t)}{dt}$ \cite{periodic}.
We consider current conservation in each node:
$ 
\sum_{\mu} I_{\mu}({\bf r})- I_{\mu}({\bf r}%
-{\bf \mu})=0,$
and we fix the total current in each direction, consistently with
the periodic boundary condition, by \cite{periodic}:
$$
I^{\rm ext}_{\mu}=\frac{\hbar}{2eR_\mu}\frac{d\alpha_{\mu}}{dt} +
\frac{1}{L_xL_yL_z}
\sum_{{\bf r}}\left[  S_{\mu}({\bf r})+\eta_{\mu}({\bf r},t)\right]
$$

We integrate with a  second order Runge-Kutta stochastic 
algorithm with $\Delta t = 0.1\, \hbar^2/(4e^2R_{ab}J) $ and 
we  average over
an interval of $10^5-10^6\Delta t$. We consider system  sizes 
$48\times 48\times L_z$ with $L_z=12-32$ and anisotropy $\Gamma=40$. 
Most of the results shown are for $L_z=32$ and  for a single disorder
realization. We have considered other realizations of disorder for
some  values of $T$ and $p$ and have found similar
behavior.

\begin{figure}[ht]
\begin{center}
\includegraphics[width=0.9\linewidth]{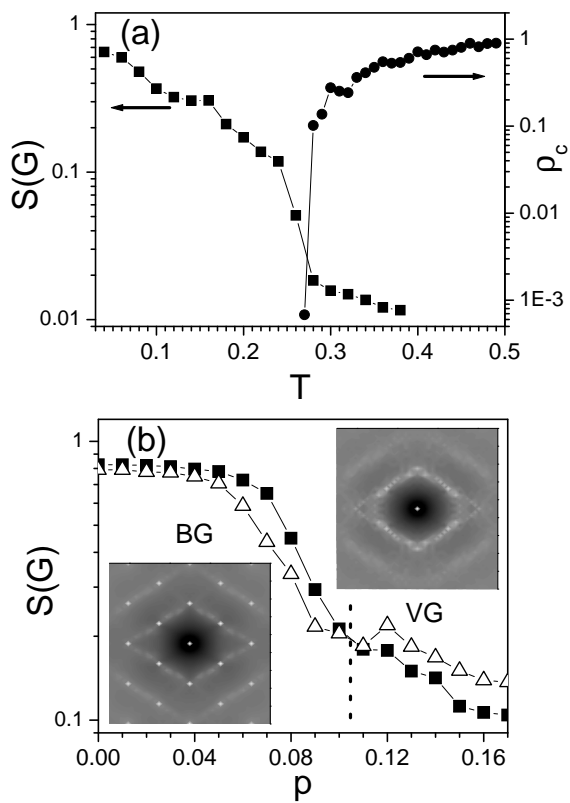}
\caption{(a) Intensity of the Bragg peak $S({\bf G})$ and $c$-axis
resistivity $\rho_c = V_z/I_z$ vs. $T$ for $p=0.01$ and
size $48\times48\times32$. $T$ is normalized by $J/k_B$.
(b) $S({\bf G})$ vs. $p$ for $T=0.1J/k_B$. $\vartriangle$:
$48\times48\times12$; $\blacksquare$: $48\times48\times32$.
Left inset: intensity plot of $S({\bf k})$ for $p=0.07$.
Right inset $S({\bf k})$ for $p=0.14$.
}
\end{center}\end{figure}

We first study the model at equilibrium (in the absence of external
currents).  At low disorder, $p=0.01$, we find a transition upon
increasing $T$ from
a vortex lattice to a vortex liquid.  We have calculated the structure  factor 
$S({\bf k})=(1/Nf^2)\langle|\sum_{{\bf r}_\perp,z} n_z({\bf r}_\perp,z)
e^{i{\bf k}.{\bf r}_\perp}|^2\rangle$, where $n_z$ is the vorticity
in the $xy$ planes and $N=L_xL_yL_z$. In Fig.~1(a) we plot
the intensity of
the structure factor, $S({\bf G})$, 
at one of the first reciprocal lattice
vectors ${\bf G}$, showing the melting
of the vortex lattice at  $T_c\approx
0.28 $. 
The resistivity is calculated by driving the system with a
very small current and obtaining voltage. We show in Fig.1(a) the
$c$-axis resistivity $\rho_c$ vs. $T$ which drops sharply at $T_c$;
we find similar behavior in $\rho_{ab}$ (not shown).
At a low temperature $T=0.1 J/k_B$, we vary disorder and study
the behavior of the lattice order parameter, $S({\bf G})$, which
is shown in Fig.~1(b) for two system sizes. 
We find that $S({\bf G})$ drops at  $p_c\approx0.11$, indicating a 
transition from the BG to the VG.
Typical structure factors $S({\bf k})$ for  $p<p_c$ and $p>p_c$ are
shown in the insets. Repeating this analysis for other $T$,
we have found that $p_c(T)$ is approximately independent
of $T$ for low temperatures. Similar phase boundary $p_c(T)$ for the
BG phase were reported in \cite{hu,teitel}. 
For very low temperatures, both \cite{hu} and \cite{teitel}
agree that there is a first-order transition from the BG phase
to the VG phase in this model. (At higher $T$ there is disagreement
on the existence of a ``vortex slush'' phase).

From now on, we focus 
on the {\it disorder driven} transition at low $T$, where
\cite{hu,teitel} agree on the phase diagram.
We fix  a low temperature $T=0.1 J/k_B$ and 
we vary the  disorder strength $p$.  In Fig.~2(a) we
show the calculated in-plane IV curves for different $p$. 
\begin{figure}[ht]
\begin{center}
\includegraphics[width=0.9\linewidth]{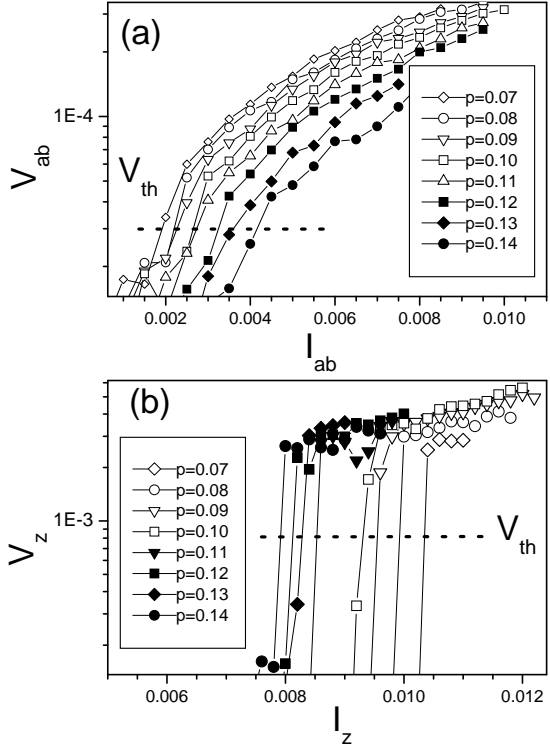}
\caption{(a) In-plane $I_{ab}$-$V_{ab}$ curves for different $p$. 
(b) $c$-axis $I_z$-$V_z$ curves. 
($I$ normalized by $I_0=2eJ/\hbar$ and $V$ by
$R_{ab}I_0$). 
}
\end{center}\end{figure}\noindent
They are obtained by applying
a uniform current along the $x$ direction, 
$I^{\rm ext}_{\mu}=I_{ab}\delta_{\mu,x}$, and calculating the 
average voltage along $x$, $V_{ab} = \langle
d\alpha_x/dt\rangle$.  In Fig.~2(b) we show the $c$-axis IV curves.
In this case we apply a uniform current along
the $\hat z$ direction (parallel to the magnetic field), 
$I^{\rm ext}_\mu = I_z\delta_{\mu,z}$, 
and we calculate the voltage $V_z = \langle d\alpha_z/dt\rangle$.
Before doing further analysis it is important to discuss the effect
of a small driving current in the BG-VG transition. 
We find that currents much smaller than the critical currents 
have a negligible effect on $p_c$. 
In Fig.~3 we show
$S({\bf G})$ vs. $p$ when the driving current in a given
direction is of the order
of the corresponding critical current. 
In the case of a $z$-direction current
$I_z=0.010 \sim I_c^z$, there
is a very small effect with a critical $p_c$ slightly smaller than
the $I=0$ case.  In the case of an in-plane current $I_{ab}=0.002\sim
I_c^{ab}$, we find that there is a first  drop of $S({\bf G})$ to a 
{\sl finite} value at a $p<p_c$. 
This corresponds to the situation when the applied current $I_{ab}$ 
equals the
critical current for that $p$. Then a second drop of $S({\bf G})$  to
the very small values of the VG phase occurs
near the critical $p_c$ but also at a slighlty smaller value. In both cases,
one can conclude that finite currents of the order of the critical
current have a small effect on the
value of $p_c$, lowering it in about $\Delta p_c \sim -0.005$.

\begin{figure}[ht]
\begin{center}
\includegraphics[width=0.9\linewidth]{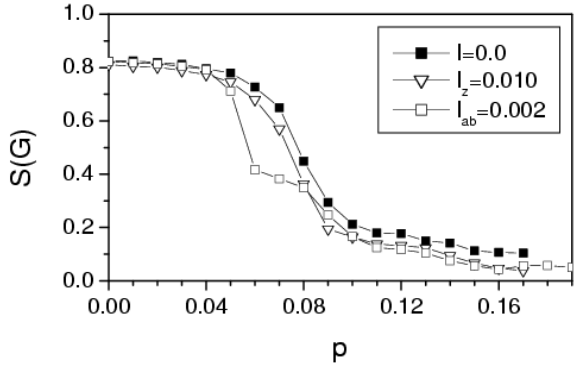}
\caption{$S({\bf G})$ vs. $p$
for finite currents, and $48\times48\times32$ size. 
}
\end{center}\end{figure}\noindent

Let us now discuss the in-plane transport at the transition. 
In Fig.2(a) we show with open
symbols the  IV curves for $p<p_c$ and with filled symbols the  IV 
curves for $p>p_c$ (where $p_c=0.11$ is the zero current critical disorder). 
We observe that in both phases the IV curves are strongly nonlinear
and that they tend to zero resistivity for low currents ($V/I
\rightarrow0$). In order to observe if there is a change when crossing
the BG-VG line, we fix a value of the current $I_{ab}$, slightly above
the critical currents, and we
calculate the voltage $V_{ab}$ as a function of disorder strength $p$.
This is shown in Fig.4(a). We see that at $p_c$ there is a 
jump down in $V_{ab}$. 
\begin{figure}[ht]
\begin{center}
\includegraphics[width=0.9\linewidth]{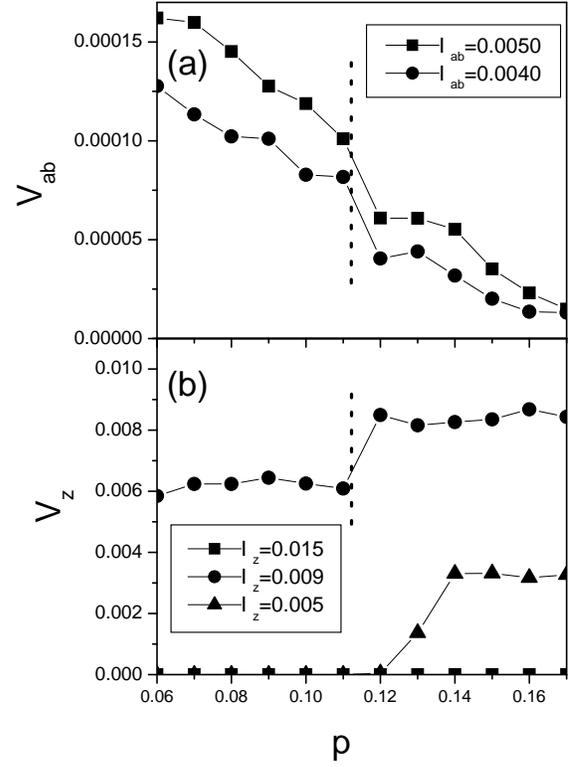}
\caption{(a) In-plane voltage $V_{ab}$ vs.  $p$.
(b) $c$-axis voltage $V_c$ vs. $p$.
(Voltages normalized by $R_{ab}I_0$, 
size is $48\times48\times32$.)
}
\end{center}\end{figure}\noindent
This jump is observed for different values of
$I_{ab}$. The location of the jump does not depend strongly on $I_{ab}$.
The experiment of \cite{safar}  finds a clear decrease of
 the differential resistance, measured for a current above the critical
 current, when increasing magnetic field at low temperatures 
 in YBaCuO samples. This
 is in agreement with our result of Fig.~4(a) for the behavior 
 of the in-plane voltage at the BG-VG transition.
One can interpret this jump down in voltage as a consequence of a jump
up in the critical current. 
It is difficult to define a critical current, since
at $T > 0$ there is always a non-zero voltage for any current.
Experimentally, the ``critical current'', $I_c$,  is obtained from
the IV curves as $V(I_c)=V_{th}$,  with the threshold voltage $V_{th}$ 
small enough. If $I_c$ is almost independent of $V_{th}$, 
then it corresponds to a crossover current separating two regimes 
in the IV curve. 
In our case, $I_c$ separates
the ohmic flux flow regime from the low current nonlinear regime 
which has $\rho_{\rm lin}(I\rightarrow0)=0$ for the two phases.
(In the case of the VG with finite screening, $\kappa<\infty$, $I_c$
would separate two ohmic regimes).
To obtain the $ab$ critical current,  $I_{ab}^c$,
we choose different values of $V_{th}$ slightly above our noise 
value of voltage. One of the $V_{th}$
used is shown in Fig.2(a) with a dashed line.
The obtained critical currents
are plotted in Fig.5(a) as a function of $p$
(we  see that different $V_{th}$  
give  smilar values of $I_c^{ab}$, within the
error bars). 
We observe in the plot that there is a  
steep increase of $I_c^{ab}$  at $p_c$.
\begin{figure}[ht]
\begin{center}
\includegraphics[width=0.9\linewidth]{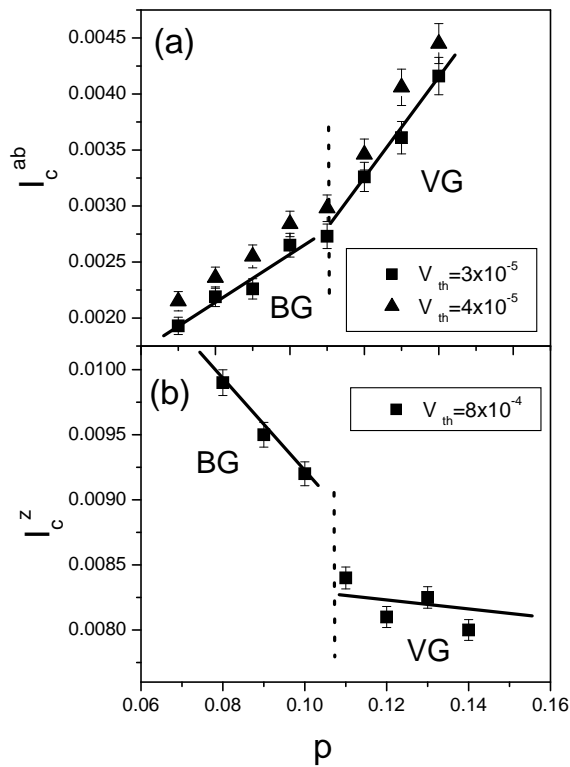}
\caption{(a) In-plane critical current $I_c^{ab}$ vs. 
$p$, for threshold voltages $V_{th}$.
(b) $c$-axis critical current $I_c^z$ vs.  
$p$. 
}
\end{center}\end{figure}\noindent
In the magnetization experiments in YBaCuO
\cite{nishizaki} the BG-VG transition line 
has been associated to the {\it onset} of the second magnetization peak, 
which has been related to an steep increase of the in-plane critical currents. 
Our result in Fig~5(a) is in good agreement with
this interpretation, particularly because 
the value of $\Gamma$ we use corresponds to YBaCuO.
The more anisotropic BiSrCaCuO samples show experimentally 
a much stronger increase in
the second magnetization peak \cite{khay}. In this very anisotropic
case the increase in critical current has also been interpreted as
a decoupling transition \cite{colson}.

The 3D XY model is very adequate to the description of the $c$-axis
transport, since the Josephson coupling between planes is treated
exactly. 
The $c$-axis current-voltage charactestics, $V_z$ vs. $I_z$,
are shown in Fig.2(b) for different values of $p$ at $T=0.1 J/k_B$. 
The  IV curves  for $p<p_c$ are plotted with open symbols and the 
IV curves for $p>p_c$
are plotted with filled symbols. We observe that in all cases the IV
curves show a well-defined critical current where the voltage drops steeply.
The  IV curves can be clearly separated in
two different sets of curves, one set for $p<p_c$ with high critical
currents and another set for $p>p_c$ with low critical currents. 
This can also be observed if for a fixed current
$I_z$  we vary $p$ and we calculate $V_z$, as we show in Fig.4(b).
For very low $I_z$ we find $V_z\approx0$ for all $p$, consistent with
both phases being superconducting. For intermediate values of $I_z$
we find that $V_z\approx0$  for $p<p_c$ with a sharp jump
to finite voltage $V_z$ at $p_c$. For values of $I_z$ above the
critical currents, there is a finite voltage $V_z$ in both phases,  
with a clear discontinuous jump up at $p_c$.
We obtain the critical currents $I_c^z$ with a voltage criterion $V_{th}$, 
which is shown in Fig.2(b) with a dashed line. 
It is obvious in this case that the values obtained for $I_c^z$ are
insesitive to the choice of $V_{th}$. The $c$-axis critical currents are
plotted as a function of disorder strength $p$ in Fig.~5(b).
We observe in the plot that $I_c^z$ has a clear discontinuous 
jump down at $p_c$.
This result is consistent with the Josephson 
plasma resonance measurements of \cite{gaifu}, 
which are sensitive to the $c$-axis Josepshon coupling, where a sharp
decrease of the plasma resonance frequency was observed at the
transition. 
Recently, the  $c$-axis Josephson critical current has been
measured experimentally in BiSrCaCuO samples varying magnetic field
at the melting transition line, the BG-VL line \cite{ooi}. A jump down of
the critical current was found, although at a   field
slightly lower than the equilibrium case. It will be interesting
if a similar experiment could be carried out 
across the BG-VG line, since our results
imply that $c$-axis transport would give a better evidence
(compared to the $ab$-plane transport)
of the first-order nature of this transition.


We acknowledge financial support from Conicet, CNEA, ANPCYT (PICT99-03-06343) 
and Fundaci\'{o}n Antorchas (Proy. 14116-147). 
A.D.H. also ackowledges support from the
Centro Latinoamericano de F\'{\i}sica.

\end{document}